\documentclass[%
 reprint,
superscriptaddress,
 amsmath,amssymb,
 aps,
pra,
]{revtex4-1}

\usepackage{graphicx}
\usepackage{dcolumn}
\usepackage{bm}
\usepackage{braket}
\usepackage{tikz}
\usepackage{booktabs}

\usepackage{algorithm}
\usepackage[noend]{algpseudocode}

\DeclareMathOperator{\e}{e}

\begin{document}


\title{Correlated Gaussian Approach to Anisotropic Resonantly Interacting Few-Body Systems}

\author{F. S. M\o ller}
\affiliation{Department of Physics and Astronomy, Aarhus University, DK-8000 Aarhus C, Denmark}
\author{D. V. Fedorov}
\affiliation{Department of Physics and Astronomy, Aarhus University, DK-8000 Aarhus C, Denmark}
\author{A. S. Jensen}
\affiliation{Department of Physics and Astronomy, Aarhus University, DK-8000 Aarhus C, Denmark}
\author{N. T. Zinner}
\email{zinner@aias.au.dk}
\affiliation{Department of Physics and Astronomy, Aarhus University, DK-8000 Aarhus C, Denmark}
\affiliation{Aarhus Institute for Advanced Studies, Aarhus University, DK-8000 Aarhus C, Denmark}

\date{\today}

\begin{abstract}
Quantum mechanical few-body systems in reduced dimensionalities can exhibit many interesting properties such as scale-invariance and universality. Analytical descriptions are often available for integer dimensionality, however, numerical approaches are necessary for addressing dimensional transitions. The Fully-Correlated Gaussian method provides a variational description of the few-body real-space wavefunction. By placing the particles in a harmonic trap, the system can be described at various degrees of anisotropy by squeezing the confinement. Through this approach, configurations of two and three identical bosons as well as heteronuclear (Cs-Cs-Li and K-K-Rb) systems are described during a continuous deformation from three to one dimension. We find that the changes in binding energies between integer dimensional cases exhibit a universal behavior akin to that seen in avoided crossings or Zeldovich rearrangement.

\end{abstract}

\maketitle

\section{Introduction}

Much can be learned from studying quantum few-body systems, as they provide the minimal example of interesting quantum phenomena, which often have great importance in the many-body context~\cite{Naidon2017,Zinner2014}. Many of these phenomena are highly dependent on the dimensionality of the system. Although analytical solutions are known in strict one-, two- and three-dimensional systems, it is much more difficult to analyze the systems during a continuous deformation from 3D $\to$ 2D and 2D $\to$ 1D. The deformation can be achieved by placing the particles in a confining potential, which is continuously squeezed in one or two directions. Realizing quasi one- or two-dimensionality experimentally often requires the system a deformation of the system, whereby the theoretical study of dimensional transitions is of interest. 

In atomic two-body systems Feshbach resonances occur, which has become one of the most important tools for controlling interactions in atomic physics~\cite{Chin2010}. Two-body systems exhibit a high degree of universality, as the s-wave scattering physics is fully determined by the scattering length $a$. For large positive $a$ in three dimensions, two particles will combine to form a weakly bound dimer state. In the resonance regime \cite{NoteResonance}, the binding energy of the dimer is given by the universal expression $E_b = -\hbar^2 / (m a^2)$, where $m$ is the atomic mass. As $1/a \to 0$, the binding energy approaches zero, and the dimer will ultimately break apart. However, in the resonance limit one will always find a bound state for any dimension less or equal to three~\cite{Petrov2001}. In fact, for any three-dimensional two-body subsystem, a strong deformation towards 2D will bind the dimer with an energy, which depends on the trap energy scale rather than the scattering length~\cite{Yamashita2014,Petrov2000,Bloch2008}. Further deformation to one dimension causes systems to exhibit new interesting properties~\cite{Olshanii1998,Peano2005}.\\
Another example of universality is found in three-body systems, where resonant bound trimer states, called Efimov states, appear with a characteristic, universal scaling in energy~\cite{Efimov1970}. While this phenomenon has been known for a while~\cite{Braaten2006,Ferlaino2010,Greene2017}, it has attracted much attention lately due to its observation in cold atoms~\cite{Kraemer2006}, Helium trimers~\cite{Kunitski2015}, and very lately in three-photon bound states~\cite{Liang2018}. For $a > 0$ the Efimov resonance corresponds to the resonant collisions between atoms and dimers~\cite{Nielsen2002,Esry1999,Esry2005,Nielsen1999,Nielsen1998}, as any scattering length beyond the binding limit of the dimer will produce an unbound atom along with the dimer. Meanwhile, for $a < 0$ one finds a bound state to three-atom continuum threshold along with the three-body resonance~\cite{Esry1999,Greene2017}. Thus, the Efimov resonance can be interpreted as a three-body generalization of the Feshbach resonance~\cite{Stoll2005}.
In three dimensions, infinitely many Efimov trimers exist in the universal limit of $1/a \to 0$. However, in two dimensions Efimov states are completely absent~\cite{Bruch1979,Nielsen1997,Lim1980}. Efimov trimers consisting of identical bosons have been predicted to only exist for dimension $d$ in the interval $2.3 \leq d \leq 3.8$~\cite{Nielsen2001,Rosa2016}. In a study~\cite{Sandoval2018} of Efimov trimers consisting of two heavy and one light particle, the collapse of Efimov states during a continuous deformation from 3D $\to$ 2D was demonstrated. This was achieved by employing a momentum space description and integral Faddeev equations~\cite{Frederico2011}.

Previous work on this subject has employed periodic boundary conditions and fitting to the trap configuration in order describe the continuous deformation of the system, whereas the Correlated Gaussian method provides a real-space calculation of the system's spectrum using the actual trap configuration.

\section{Methods}

The Correlated Gaussian Method is a variational method, which has been used to solve quantum-mechanical few-body problems in molecular, atomic and nuclear physics~\cite{Bubin2012,Mitroy2013,Sharkey2011,Daily2014,Tung2014}. The method has attained its popularity from the ease of calculating matrix elements in a Gaussian basis~\cite{Varga1995,Suzuki1998}. Many matrix elements and their corresponding gradients are fully analytic, whereby numerical optimizations can be carried out with a high degree of accuracy~\cite{Sorensen2005,Fedorov2017}.

\subsection{Fully Correlated Gaussian Method}
The standard procedure when employing the correlated Gaussian method is utilizing "isotropic" Gaussians, which do not distinguish between directions for the individual particles. This approach is very efficient, as only few optimization parameters are needed. The lack of explicit directional bias is compensated by shifting the positions of the Gaussians, thus enabling accurate approximations of most wavefunctions. However, this approach is not feasible in squeezing problems, since the size of the Gaussians is limited by the width of any confining potential. As the potential is squeezed tightly in one direction, the number of isotropic Gaussians needed to cover the entire wavefunction will increase significantly.
Hence, an alternative method is presented here, where fully correlated Gaussians are employed, which do distinguish directions at the cost of additional variational parameters. 
For a system of $N$ particles with coordinates $\boldsymbol{r} = (\vec{r}_1 , \vec{r}_2 , \ldots , \vec{r}_N)^{\mathrm{T}} = (r_1 , r_2 , \ldots , r_{3 N})^{\mathrm{T}}$, the fully correlated Gaussians have the form
\begin{align}
 \braket{ \boldsymbol{r} | g }  &\equiv  \exp \left( - \sum_{i < j = 1}^{3 \cdot N} a_{ij} (r_i - r_j)^2 + \sum_{i=1}^{3 \cdot N} s_i \cdot r_i \right)  \nonumber \\
  &=   \exp \left( - \sum_{i < j = 1}^{3 \cdot N} a_{ij} \boldsymbol{r}^{\mathrm{T}} w_{ij} w_{ij}^{\mathrm{T}} \boldsymbol{r} + \sum_{i=1}^{3 \cdot N} s_i \cdot r_i \right) \nonumber \\
 &=  e^{-\boldsymbol{r}^{\mathrm{T}} A \boldsymbol{r} + \boldsymbol{s}^{\mathrm{T}} \boldsymbol{r} } \; ,
\end{align}
where $\boldsymbol{s}$ is a column of variational shifts, $A$ is a positive-definite matrix containing the variational parameters $a_{ij} > 0$, and $w_{ij}$ is a size $3 N$ column vector of zeros with the exception of elements $i$ and $j$ being 1 and -1 respectively. 
This particular form is chosen to ensure the positive-definiteness of $A$, as the matrix is constructed as a sum of positive-definite rank-1 matrices $a_{ij}  w_{ij} w_{ij}^{\mathrm{T}}$. 
The trial wavefunction of the few-body wavefunction, $\ket{\psi}$, is written as a linear combination of $K$ Gaussians
\begin{equation}
	\ket{\psi} = \sum_{i = 1}^{K} c_i \ket{g_i}
\end{equation}
Inserting this expression into the Schr{\"o}dinger equation
\begin{equation}
	\hat{H} \sum_{i=1}^{K} c_{i} \ket{g_i} = E \sum_{i=1}^{K} c_{i} \ket{g_i} 
\end{equation}
and multiplying from the left with $\bra{g_j}$ yields
\begin{equation}
	\sum_{i=1}^{K} c_{i} \bra{g_j} \hat{H} \ket{g_i} = E \sum_{i=1}^{K} c_{i} \braket{g_j | g_i} \; .
\end{equation}
This expression can be formulated as a generalized eigenvalue problem
\begin{equation}
	\mathcal{H} \boldsymbol{c} = E \mathcal{B} \boldsymbol{c} \; ,
	\label{eq:genEigenProb}
\end{equation}
where $\boldsymbol{c}$ is the column of linear parameters, $c_i$, while $\mathcal{H}_{j,i} \equiv \bra{g_j} \hat{H} \ket{g_i}$ and $\mathcal{B}_{j,i} \equiv \braket{g_j | g_i}$. A downside of utilizing a basis of Gaussians is the non-orthogonality of the functions resulting in a non-identity overlap-matrix, $\mathcal{B}$. The generalized eigenvalue problem can be turned into a regular eigenvalue problem through a Cholesky decomposition, and the symmetry of the matrices $\mathcal{H}$ and $\mathcal{B}$ can be exploited for obtaining solutions faster. 
While the linear parameters, $c_i$, together with the energy spectrum are found by solving eq. \eqref{eq:genEigenProb}, the non-linear parameters, $a_{ij}$ and $s_i$, are found using the Nelder-Mead optimization algorithm~\cite{Nelder1965,Nocedal2006}.

\subsection{Jacobi Coordinates}
It is convenient to use a set a Jacobi coordinates, $\boldsymbol{x} = (\vec{x}_1 , \ldots, \vec{x}_{N-1})^{\mathrm{T}}$, rather than relative distance vectors, $(\vec{r}_i - \vec{r}_j)$ \cite{Varga1995}. The Jacobi coordinates separate the center-of-mass coordinates from the rest, which enables a description of the internal dynamics of the systems. Furthermore, disregarding the center-of-mass coordinates reduces the amount of parameters needed.
Consider the transformation of the coordinate vector, $\boldsymbol{r}$, and its corresponding gradient operator 
\begin{equation}
	\boldsymbol{x} = \mathcal{U} \boldsymbol{r} \quad , \quad \hat{\boldsymbol{\nabla}} = \mathcal{U}^{\mathrm{T}} \hat{\boldsymbol{\nabla}}_x \; ,
\end{equation}
where $\mathcal{U}$ is the transformation matrix and $\hat{\boldsymbol{\nabla}} = \left( \hat{\nabla}_1 , \ldots , \hat{\nabla}_N \right)^{\mathrm{T}}$. 	
In order to produce Jacobian coordinates, a possible form of the transformation matrix is \cite{Mitroy2013}
\begin{equation}
 \mathcal{U} \: = \: \begin{pmatrix}
1 & -1 & 0 & \hdots & 0 \\
\frac{m_1}{m_1 + m_2} & \frac{m_2}{m_1 + m_2} & -1 & \hdots & 0 \\
\vdots & \vdots & \vdots & \ddots & \vdots \\
\frac{m_1}{m_1 + \ldots + m_N} & \frac{m_2}{m_1  + \ldots + m_N} & \hdots & \hdots & \frac{m_N}{m_1  + \ldots + m_N}
\end{pmatrix} \; .
	\label{eq:JacobiTransMatrix}
\end{equation}
Although this transformation matrix is not unitary, its inverse can be written explicitly \cite{Suzuki1998}.
In general, in the presence of external fields it is not always possible to separate out the contributions from the center-of-mass. However, if one assumes the same external harmonic one-body potential on each particle, the internal part of the system can be treated separately from the center-of-mass
\begin{equation}
	\hat{T} = \hat{T}^{\mathrm{int}} + \hat{T}^{\mathrm{CM}} \quad , \quad \hat{V}_{\mathrm{HO}} = \hat{V}_{\mathrm{HO}}^{\mathrm{int}} + \hat{V}_{\mathrm{HO}}^{\mathrm{CM}} \; .
\end{equation}
Thus, the total wavefunction of the system can be factorized as  
\begin{equation}
	\Psi \left( \vec{r}_1 , \vec{r}_2 , \ldots , \vec{r}_N \right) = \Phi \left( \vec{x}_1 , \vec{x}_2 , \ldots , \vec{x}_{N-1} \right) \phi( \vec{x}_N ) \; ,
\end{equation}
where the wavefunction dependent on the center-of-mass coordinate, $\vec{x}_N$, can be completely neglected.
Therefore, considering only harmonic confining potentials as external fields, the few-body Hamiltonian considered in Jacobi coordinates reads 
\begin{equation}
	\hat{H} = - \frac{\hbar^2}{2} \hat{\boldsymbol{\nabla}}_{x}^{\mathrm{T}} \Lambda \hat{\boldsymbol{\nabla}}_x + \boldsymbol{x}^{\mathrm{T}} \Omega \boldsymbol{x} + \sum_{i < j}^{N} V_{ij}(\boldsymbol{x}) + \hat{H}^{\mathrm{CM}} \; ,
\end{equation}
where $V_{ij}(\boldsymbol{x})$ is an arbitrary two-body potential. The matrix $\Lambda$ has elements of the form
\begin{equation}
	\Lambda_{kj} = \sum_{i = 1}^{N} \frac{\mathcal{U}_{ki} \mathcal{U}_{ji}}{m_i} \; ,
	\label{eq:LambdaElements}
\end{equation}
which can be derived from the separation of the kinetic operator. The elements \eqref{eq:LambdaElements} are general, however, in the scenario of the transformation matrix being that of eq. \eqref{eq:JacobiTransMatrix}, the elements reduce to $\Lambda_{kj} = \mu_{k}^{-1} \delta_{kj}$, where $\mu_k = \frac{m_{k+1} \left( \sum_{i= 1}^{k} m_{i} \right)}{\sum_{i= 1}^{k+1} m_{i}}$ is the $k$'th reduced mass.
Following this, the elements of the harmonic oscillator matrix $\Omega$ can be written as
\begin{equation}
	\Omega_{kj} = \frac{1}{2} \mu_{k} \delta_{kj}  \omega^2  \; .
\end{equation}
In the case of full correlation the frequency $\omega$ is given by a vector, $\boldsymbol{\omega} = (\omega_x , \omega_y , \omega_z)$.

\section{Results}

We now discuss our results for different problems involving few-body systems in squeezed geometries. In natural progression, we start with a two-body system of identical bosons, then advance to three identical bosons and lastly discuss two identical bosons and a third particle. All the calculations are performed using a basis of fully correlated Gaussians optimized with the Nelder-Mead algorithm. The energies calculated are in units of $\hbar ^2 / 2 m r^2$, where $m$ is the mass scale set by the lightest particle in the system, and $r$ is the length scale set by the range of the two-body potentials, $r_0$. 

\subsection{Universality in AA-system}
First, we consider a two-body system with particles of equal mass. For this system three different interaction potentials are explored,
\begin{align}
	V_{i j}^{(1)} (w_{i j}^{\mathrm{T}} \boldsymbol{r}) &= - S_1 \e ^{- \boldsymbol{r}^{\mathrm{T}} w_{i j} w_{i j}^{\mathrm{T}} \boldsymbol{r} / r_{0}^2 } \; ,	\label{eq:Vij1} \\
	V_{i j}^{(2)} (w_{i j}^{\mathrm{T}} \boldsymbol{r}) &= 2 S_2 \e ^{- \boldsymbol{r}^{\mathrm{T}} w_{i j} w_{i j}^{\mathrm{T}} \boldsymbol{r} /(r_{0} /2)^2 } - S_2 \e ^{- \boldsymbol{r}^{\mathrm{T}} w_{i j} w_{i j}^{\mathrm{T}} \boldsymbol{r} /r_{0}^2 }	, \label{eq:Vij2} \\
	V_{i j}^{(3)} (w_{i j}^{\mathrm{T}} \boldsymbol{r}) &= - S_3 \e ^{- r_{0} w_{i j}^{\mathrm{T}} \boldsymbol{r} } \; .	\label{eq:Vij3}
\end{align}
The interaction strengths are tuned to resonant scattering in three dimensions, whereby $|a_{\mathrm{3D}}| = \infty$. These simple potentials are used to model short range interactions. In the absence of the external trapping confinement, zero-range interactions with large scattering length can be modeled by setting $r_0$ small~\cite{Nielsen2001}.
The resonant strengths along with the corresponding effective ranges of the potentials are shown in table \ref{tab:TunedStrengths}.
\begin{table}[!h]
\centering
\begin{tabular}{l|lll}
   & $V_{ij}^{(1)}$ & $V_{ij}^{(2)}$ & $V_{ij}^{(3)}$ \\
[1mm] \hline \\[-2mm]
$2 \mu r_{0}^{2} S/\hbar^2$ & 2.684          & 4.188          & 1.4456         \\
$R_e$                       & 1.4394         & 1.7061         & 3.5483        
\end{tabular}
\caption{Interaction strengths tuned for infinite scattering length along with corresponding effective ranges in units of $r_0$.}
\label{tab:TunedStrengths}
\end{table}
The two particles are subjected to the same single-particle harmonic trapping potential in one direction $\frac{1}{2} m \omega_{x}^2 x^2$, while the remaining directions remain free. Starting at $b_{x} = \sqrt{\hbar / \mu \omega_x} \gg 1$, the system is continuously squeezed in the x-direction by letting $b_x \to 0$.
\begin{figure}[!t]
	\centering
	\includegraphics[width=1.1\columnwidth]{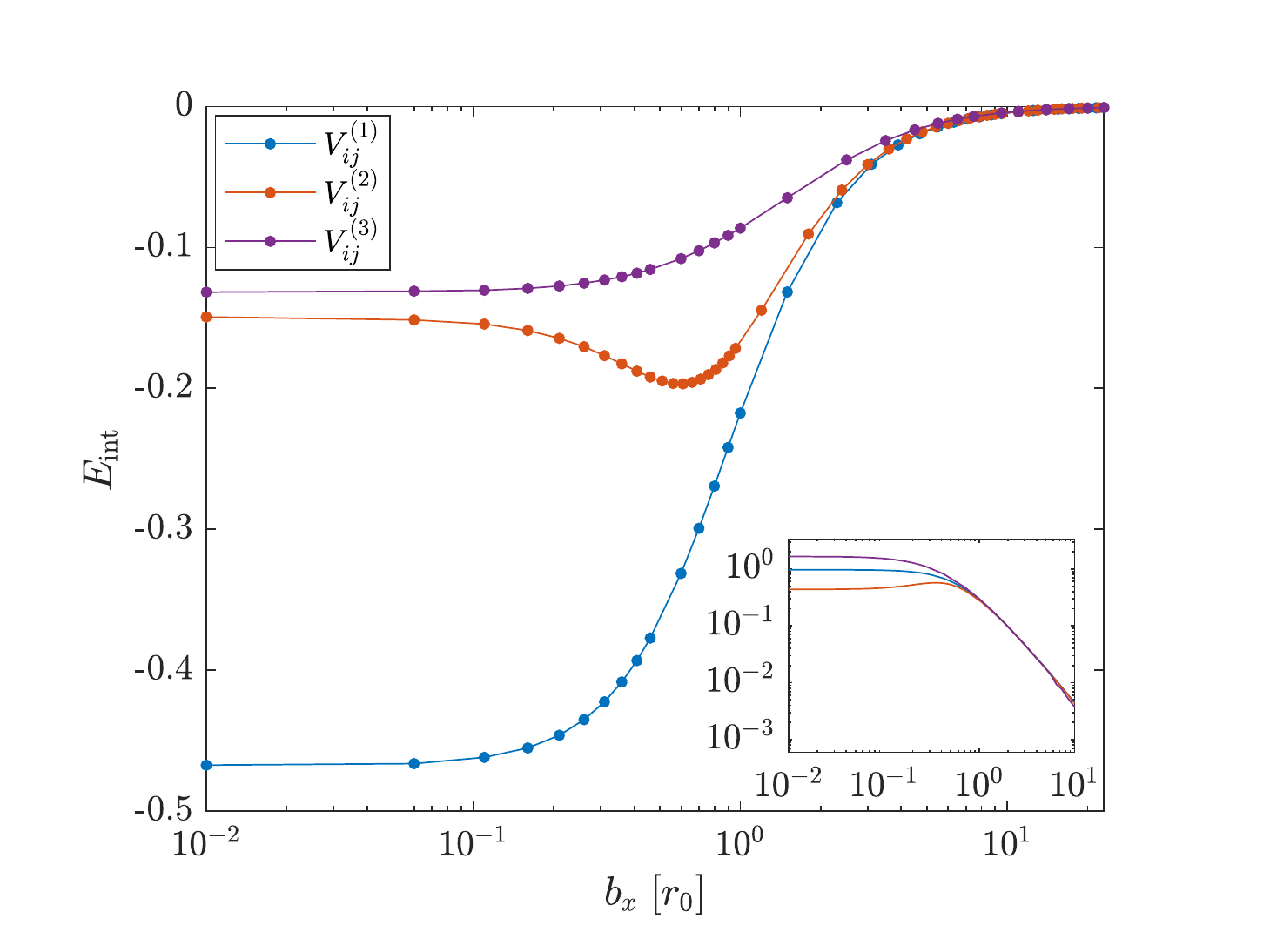} 
	\caption{Ground state energies of AA-system with two-body potentials \eqref{eq:Vij1}, \eqref{eq:Vij2} and \eqref{eq:Vij3} as function of the squeezing parameter $b_x$. The ground state harmonic oscillator energy has been subtracted. The inset shows the binding energy as a function of $b_x$, where axis have been scaled with the effective range of the two-body potentials.}
	\label{fig:AAsqueeze} 
\end{figure}
\begin{figure*}[!t]
	\centering
	\includegraphics[width=0.8\textwidth]{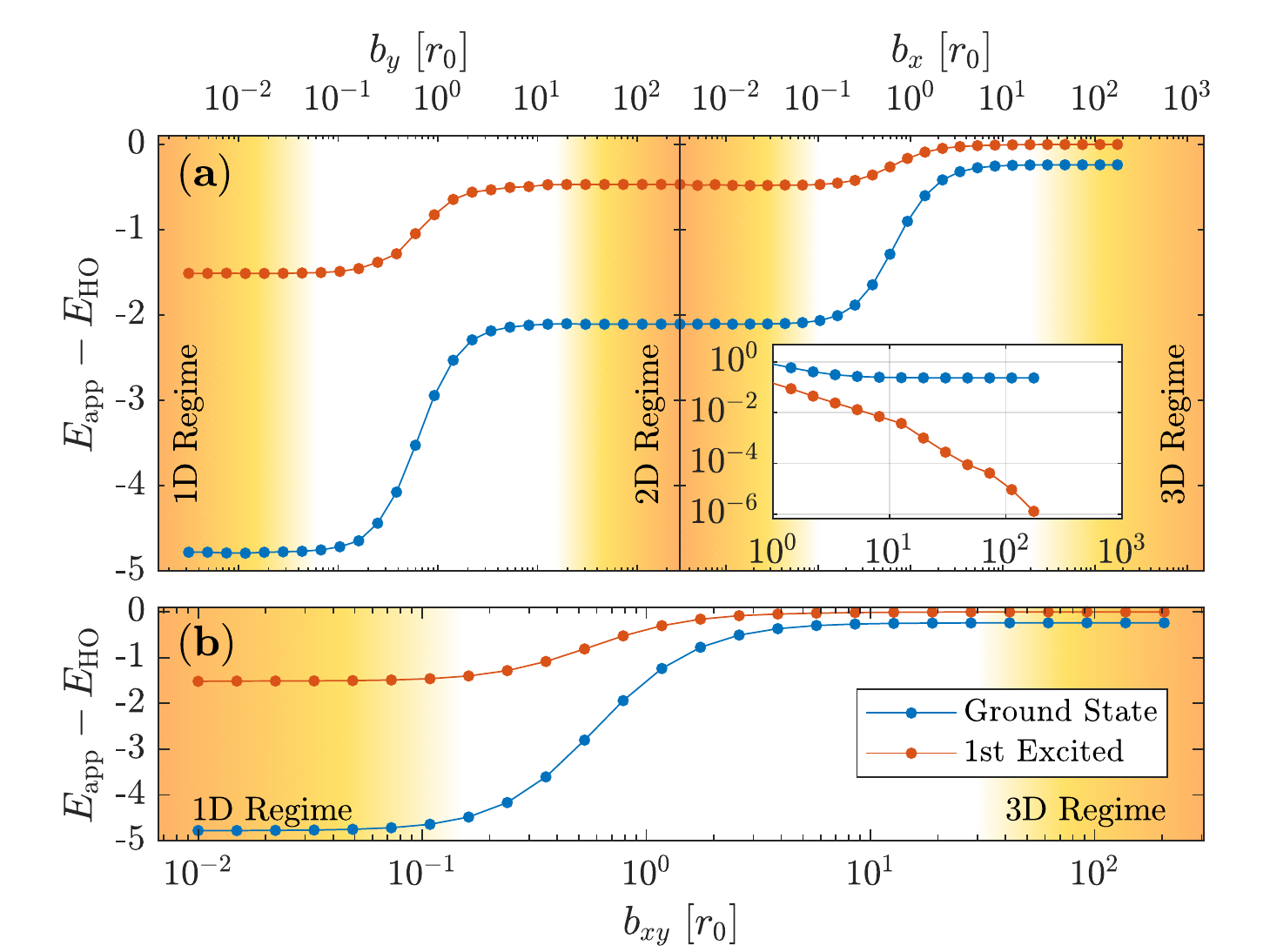} 
	\caption{Three-body spectrum of AAA-system for various widths of the harmonic oscillator trap. \textbf{(a)} the confinement is squeezed in the $x$- and $y$-direction separately. The inset displays the binding energy in the 3D regime as function of $b_x$. \textbf{(b)} simultaneous squeezing in $x$- and $y$-direction.}
	\label{fig:AAAsqueeze} 
\end{figure*}
The internal energy of the AA system as function of $b_x$ can be seen in fig. \ref{fig:AAsqueeze}. To obtain the internal energy, the harmonic ground state energy has been subtracted, as this would otherwise overshadow the subtle change in the energy spectrum due to the squeezing. As expected, the system evolves from the zero-energy state to a bound dimer. Initially, the evolution of the system remains constant until $b_x \sim r_0$, where the confinement is strongly felt by the particles, and the energy evolves quickly towards the 2D limit. As the system is squeezed, the different features of the potentials do not become apparent before the system approaches the 2D limit. When $|a_{\mathrm{3D}}|$ greatly exceeds the range of the two-body potential, the details of the potential become irrelevant due to the long, exponential tail of the wavefunction. However, as the system is squeezed the scattering lengths diverges from its 3D limit and the features of the potentials become apparent.
To illustrate this, the axis of fig. \ref{fig:AAsqueeze} has been scaled with the effective ranges shown in table \ref{tab:TunedStrengths}, resulting in the inset of the figure. Until the width of the confinement trap reaches the effective range $b_x / R_e \sim 1$ the system evolves universally.

\subsection{AAA-system}
\begin{figure*}[!t]
	\centering
	\includegraphics[width=0.8\textwidth]{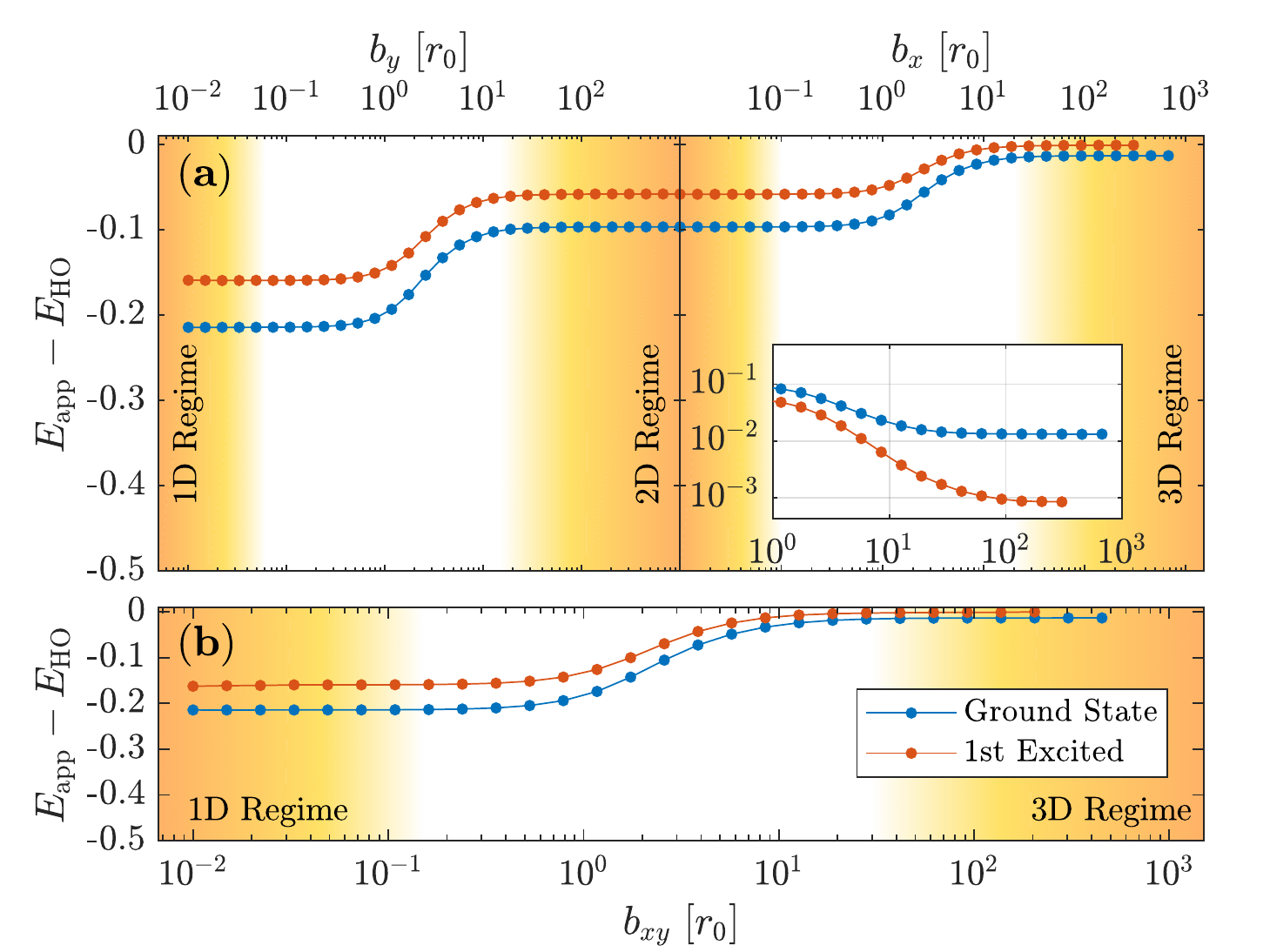} 
	\caption{Three-body spectrum of AAB-system ($m_B / m_A = 6/133$) for various widths of the harmonic oscillator trap. \textbf{(a)} the confinement is squeezed in the $x$- and $y$-direction separately. The inset displays the binding energy in the 3D regime as function of $b_x$. \textbf{(b)} simultaneous squeezing in $x$- and $y$-direction.}
	\label{fig:AABsqueeze} 
\end{figure*}
Next, we consider a systems of three particles with equal mass and infinite scattering lengths. The particles are interacting through the single Gaussian interactive potential \eqref{eq:Vij1}. The particles are subjected to two harmonic confinements in the x- and y-direction respectively. Thus, the total trapping potential reads
\begin{equation}
	V_{trap} = \frac{1}{2} m \omega_{x}^2 x^2 + \frac{1}{2} m \omega_{y}^2 y^2 \; ,
\end{equation}
and is experiences by all the particles. Initially this confinement is very weak. First, the system is continuously squeezed in the x-direction by letting $b_x \to 0$ or $\omega_x \to \infty$. As the system reaches the limit 2D-limit, the width of the trap in the y-direction is decreased until the system finally becomes effectively one-dimensional.
Figure \ref{fig:AAAsqueeze}(a) displays the results from calculating selected energy levels of the system at various trap configurations. 
Even though the AA subsystems have zero binding energy in the three-dimensional limit, the binding energy of the three-body system is positive and finite due to the collective interactions.
The transition from 3D $\to$ 2D is very similar to the two-particle case, as the spectrum changes slowly until $b_x \sim r_0$. The inset of figure \ref{fig:AAAsqueeze} shows the binding energy in the three-dimensional limit as a function of the trap width, $b_x$. Notice how the higher excited states are affected by the squeezing first, as these states have a larger spatial extension, thus making them more sensitive to confinements. The Fully Correlated Gaussian method struggles describing the wavefunction in the three-dimensional limit, which is due to the difficulty of fitting the long exponential tail of the wavefunction. Similarly, calculating excited states become increasingly difficult due to their larger spatial extension and in turn longer range tail. In Appendix \ref{chap:disc} a series of possible solutions and extensions to the method are proposed.

The evolution of the spectrum during the following transition from 2D $\to$ 1D proceeds similar to the previous transition. While the different states respond in a similar manner to the squeezing, the magnitude of the change in internal energy is very state dependent. This illustrates the difference in dynamics between the various dimensionalities.\\
Instead of deforming the confining potentials one at a time, another possibility is performing a simultaneous squeezing in the x- and y-direction. The transition to the one-dimensional limit is of great interest, as three-body, quasi-one-dimensional systems exhibit universality in both the bosonic~\cite{Peano2005,Mora2005} and fermionic case~\cite{Mora2004,Gogolin2005}. 
The resulting spectrum of the direct transition to the 1D limit can be seen in figure \ref{fig:AAAsqueeze}(b). While the one- and three-dimensional limits are the same as for the separate squeezing, the system is never confined to two dimensions, as the last dimension remain free. Thus, the system does not exhibit quasi 2D dynamics, which is apparent from the lack of a plateau in the spectrum otherwise present in figure \ref{fig:AAAsqueeze}(a). Instead, the simultaneous squeezing causes a rapid transition from 3D $\to$ 1D, which takes place for trapping widths of $b_{xy} \sim 0.1 - 10$.

Comparing the simultaneous and sequential squeezing reveals that the transition to a lower dimensionality occurs at roughly the same trap-widths, no matter which trap configuration the system is being confined to. This sort of universal behavior is similar to what was observed for the AA-system in figure \ref{fig:AAsqueeze}. The rapid change of the binding energy occurs when the trap widths becomes comparable to the length scale of the state. Thus, the microscopic details of the state are irrelevant for the confinement leading to the observed universality.
This universal behavior is akin to the phenomenon of Zeldovich rearrangement, which was originally considered for electrons in magnetic fields~\cite{Zeldovich1960} and is an effect that occurs for systems where there is a strong competition between long-range and short-range interactions and/or potential fields. The analogy to the systems we consider here is that we have short-range interactions and trapping potentials that exert their influence at long-range. Hence, one may see a similar competition of short- and long-range effects and get the rearrangment of levels characterized by sudden large changes in level energies where a lower lying level pushes a level above upward and takes it place.

\subsection{AAB-system}
\begin{figure*}[!t]
	\centering
	\includegraphics[width=0.8\textwidth]{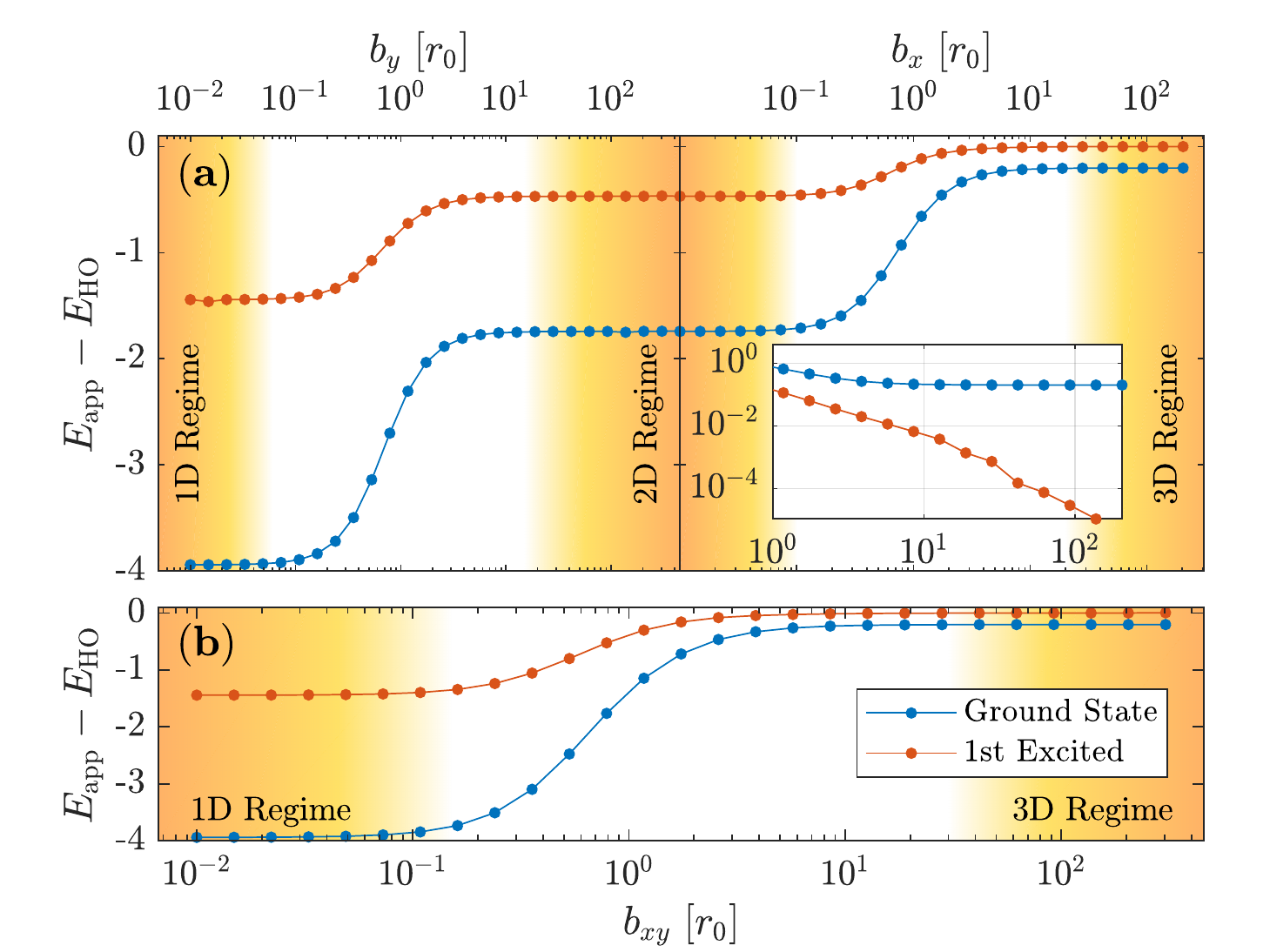} 
	\caption{Three-body spectrum of ABB-system ($m_B / m_A = 41/87$) for various widths of the harmonic oscillator trap. \textbf{(a)} the confinement is squeezed in the $x$- and $y$-direction separately. The inset displays the binding energy in the 3D regime as function of $b_x$. \textbf{(b)} simultaneous squeezing in $x$- and $y$-direction.}
	\label{fig:ABBsqueeze} 
\end{figure*}
Another interesting candidate for observing Efimov physics are heteronuclear systems~\cite{Rosa2018,Mikkelsen2015,Petrov2015}, especially systems consisting of two heavy particles and a single light particle. The high ratio in mass results in a large reduced mass, which in turn causes a smaller separation of the Efimov levels~\cite{Jensen2003}. The system examined in this report is meant to emulate that in \cite{Sandoval2018} and provide a calculation of the spectrum in real-space. The system in \cite{Sandoval2018} is meant to represent a mixture of Li-6 and Cs-133 with mass ratio $m_B / m_A = 6/133$, as these systems have shown great experimental potential for studying Efimov physics~\cite{Pires2014,Tung2014scaling,Ulmanis2016}. Furthermore, the interaction between the two heavy particles was turned off. 

The spectrum calculated in \cite{Sandoval2018} was scaled with the 3D dimer energy. As this energy tends towards zero for $|a_{\mathrm{3D}}| = \infty$, a finite, positive scattering length was chosen. No exact scattering length was provided in \cite{Sandoval2018}, hence an interaction strength of $2 \mu r_{0}^{2} S_1/\hbar^2 = 2.8$ was chosen for the calculation presented in this report. Solving the one-dimensional Schr{\"o}dinger equation for the scattering potential resulted in a scattering length of $ a = 26.94 r_0 $ corresponding to a dimer energy of $ E_{\mathrm{3D}}^{(2)}  = 7.200\mathrm{e}-4$.

Similarly to the AAA-system, a sequential 3D $\to$ 2D $\to$ 1D transition was achieved by squeezing confinement potentials in the x- and y-direction one at a time. The results can be seen in figure \ref{fig:AABsqueeze}(a). Due to the large reduced mass, the spacing between the energy levels is much less than for the AAA-system. Furthermore, the heavy mass of the particles causes a smaller spatial extension of the wavefunctions. Therefore, calculating the spectrum in the three-dimensional limit is generally easier for the AAB-system, which is illustrated in the inset of figure \ref{fig:AABsqueeze}, where the first excited state converges nicely.\\

Similarly to the AAA-system, a simultaneous squeeze in both the $x$- and $y$-direction of the AAB-system was performed. The results can be seen in figure \ref{fig:AABsqueeze}(b).
Again, one observes the same energy reached in the 1D limit as when squeezing in the two directions one at a time. Compared to the AAA-system, the transition rate of the heavily mass-imbalanced AAB system is faster. This is especially apparent when comparing the two simultaneous squeezes, where the AAB system achieves quasi one-dimensional dynamics after a change in confinement width between $b_{xy} \sim 1 - 10$. Nevertheless, the system still exhibits the same universal behavior when squeezing in one or two directions, which was observed for the case of identical particles.

\subsection{ABB-system}
Lastly, another type of heteronuclear system experimentally investigated for occurrences of Efimov phenomena are K-Rb mixtures~\cite{Barontini2009,Helfrich2010,Wacker2015}. Here, a three-body system consisting of a single Rb-87 atom with two K-41 atoms is considered with a mass ration $m_B / m_A = 41/87$~\cite{Wacker2016}. The system features resonant two-body interactions described by eq. \eqref{eq:Vij1} between all particles.\\
The results of the calculations can be seen in figure \ref{fig:ABBsqueeze}. The ABB-system behaves very similarly to the AAA-system, as their mass ratio is comparable. It should be stressed again that the energy is scaled according to the lightest particle of the system. Nevertheless, the universal behavior during both the sequential (figure \ref{fig:ABBsqueeze}.(a)) and simultaneous (figure \ref{fig:ABBsqueeze}.(b)) confinement squeezing is similar to the two previous cases.

\section{Conclusion and Outlook}
Lately there has been an experimental push towards systems of lower dimensionality, as these systems often exhibit novel phenomena and universality~\cite{Boettcher2016,Murthy2018,Hung2011,Feld2011}. A proper understanding of the underlying few-body physics is often necessary for the description of many-body phenomena~\cite{Zinner2014,Bloch2008}. The Fully-Correlated Gaussian method provides real-space solutions of highly anisotropic few-body systems, which is illustrated in this paper by calculating the spectrum of both homo- and heteronuclear systems at various degrees of transversal confinement. For systems of two identical bosons, the method showed universality in the evolution of spectra during a transition from 3D $\to$ 2D for different interactions. Following this, several three-body systems were investigated during a transition from 3D $\to$ 1D, which was achieved through both a sequential and a simultaneous squeezing of the confining harmonic potential. The difference between the two approaches was illustrated by the onset of quasi-2D dynamics for the sequential squeeze, which was apparent from the plateau in the spectrum. 

Furthermore, the transition between different dimensionalities occurs rapidly during changes of trapping widths, whereas the system is basically unaffected by any previous and subsequent squeezing. The regions of trap widths, in which the system is affected by the squeezing, display a high degree of universality, as they are independent of the current dimensionality and number of squeezing directions. 

While the Fully-Correlated Gaussian method accurately describes lower excited states, the method struggles as the spatial extension of the system grows larger, due to the difficulty of capturing the behavior of the long exponential tail of the wavefunction. Therefore, further extensions  to the method (discussed in Appendix \ref{chap:disc}) are needed, if Efimov states are to be modeled.

\appendix

\section{Matrix Elements} \label{chap:MatrixElem}
As mentioned earlier, the great advantage of using correlated Gaussians as basis function is the analytical matrix elements. Here, the matrix elements used for the calculations are presented. The expressions originate from \cite{Fedorov2017}. The elements have been modified to account for fully correlated Gaussians of the form
\begin{equation}
 \braket{ \boldsymbol{r} | g }  =  e^{-\boldsymbol{r}^{\mathrm{T}} A \boldsymbol{r} + \boldsymbol{s}^{\mathrm{T}} \boldsymbol{r} } \; .
\end{equation}
First, for a cleaner notation the following expressions are used
\begin{align}
	B &= A + A' \\
\boldsymbol{v} &= \boldsymbol{s} + \boldsymbol{s}' \\
\boldsymbol{u} &= \frac{1}{2} B^{-1} \boldsymbol{v} \; .
\end{align}
As Gaussians are not orthogonal, they have a finite overlap. The overlap matrix element, $M$, is given by
\begin{equation}
	\braket{ g' |  g} = \e ^{\frac{1}{4} \boldsymbol{v}^{\mathrm{T}} B^{-1} \boldsymbol{v}}  \frac{\pi ^{3 n /2}}{\sqrt{\mathrm{det}(B)}} \equiv M  \; ,
\end{equation}
where $n = N-1$ is the reduced number of bodies after removing the center-of-mass.
Two common matrix element needed for computations are
\begin{equation}
	\braket{ g' | \boldsymbol{r} | g} =  \frac{\partial}{\partial \boldsymbol{v}^{\mathrm{T}}}  M = \boldsymbol{u} M \; ,
\end{equation}
and
\begin{align}
	\braket{ g' | \boldsymbol{r}^{\mathrm{T}} F \boldsymbol{r} | g} &= \left( \frac{\partial}{\partial \boldsymbol{v}} F \frac{\partial}{\partial \boldsymbol{v}^{\mathrm{T}}} \right) M \nonumber \\
	&= \left( \frac{1}{2} \mathrm{trace} \left( F B^{-1} \right) + \boldsymbol{u}^{\mathrm{T}} F \boldsymbol{u}  \right) M	 \; . \label{eq:xFx}
\end{align}
From these elements the kinetic matrix element can be derived as
\begin{align}
	\braket{ g' | - \hat{\boldsymbol{\nabla}}_{x}^{\mathrm{T}} \Lambda \hat{\boldsymbol{\nabla}}_x | g} &= \Big( 2 \mathrm{trace} \left( A' \Lambda A B^{-1} \right) \nonumber \\
	&\quad + \left( \boldsymbol{s}' - 2 A' \boldsymbol{u} \right)^{\mathrm{T}} \Lambda \left( \boldsymbol{s} - 2 A \boldsymbol{u} \right) \Big) M \; ,
\end{align}
where the matrix $\Lambda$ is given by
\begin{equation}
	\Lambda_{kj} = \sum_{i = 1}^{N} \frac{\mathcal{U}_{ki} \mathcal{U}_{ji}}{m_i} \; .
\end{equation}
The two-body potentials used in this report are Gaussian shaped, whereby the derivation of the overlap can be reused. Thus, the matrix of the single Gauss potential of eq. \eqref{eq:Vij1} reads
\begin{align}
	\Braket{V_{ij}^{(1)}} &=  \Braket{ g' | \e ^{- \gamma \boldsymbol{r}^{\mathrm{T}} w_{i j} w_{i j}^{\mathrm{T}} \boldsymbol{r}  }  |  g} \nonumber \\
	&= \e ^{\frac{1}{4} \boldsymbol{v}^{\mathrm{T}} \tilde{B}^{-1} \boldsymbol{v}}  \frac{\pi ^{3 n /2}}{\sqrt{\mathrm{det}(\tilde{B})}} \equiv \tilde{M}  \; ,
\end{align}
where the matrix $\tilde{B}$ is a rank-1 update of the matrix $B$, such that
\begin{equation}
	\tilde{B} = B + \gamma w_{i j} w_{i j}^{\mathrm{T}} \; .
\end{equation}
Note, in the case of full correlation $B$ must be updated for each of the coordinates. Thus, three rank-1 updates must be made. Alternative an update matrix can be constructed as the sum of the three updated
\begin{equation}
	W_{i j} = \sum_{c = x, y, z} w_{i j \; (c)} w_{i j \; (c)}^{\mathrm{T}} \; .
\end{equation}
Lastly, the matrix element of the harmonic oscillator can be expressed using eq. \eqref{eq:xFx}. Thereby the matrix element reads
\begin{equation}
	\braket{ g' | \boldsymbol{r}^{\mathrm{T}} \Omega \boldsymbol{r} | g} = \left( \frac{1}{2} \mathrm{trace} \left( \Omega B^{-1} \right) + \boldsymbol{u}^{\mathrm{T}} \Omega \boldsymbol{u}  \right) M \; ,
\end{equation} 
where the matrix $\Omega$ is given by
\begin{equation}
	\Omega_{kj \; (c)} = \frac{1}{2} \mu_{k} \delta_{kj}  \omega_{c}^2  \; .
\end{equation}
Again, one should note the how all coordinates $c = x, y, z$ must be taken into account.

\section{Extensions to the Correlated Gaussian Method} \label{chap:disc}
The calculations presented here were performed using a fixed basis size chosen at the start of each run. The basis functions were optimized one at a time through several sweeps of the entire basis. A major disadvantage of this approach is that the basis size needed to accurately describe the wavefunction is unknown. Therefore, one often has perform the same calculation for a series of basis sizes in order to gauge the correct amount of Gaussians needed for the description.\\
An alternative method is adding an optimized function to the basis one at time, until the energy no longer improves. This approach is more flexible, as no knowledge of the required basis size is needed. Furthermore, the method has a well defined convergence criteria, as one simply stops adding more functions to basis once no improvement in the energy occurs. However, this approach may introduce artificial minima to the optimization landscape, as functions added to the basis are optimized with regards to the current configuration of Gaussians rather than the underlying wavefunction. Furthermore, the first couple of functions added to the basis are often a poor description of the wavefunction, whereby this approach often requires a larger basis in the end.\\
Another approach worth examining is combining the two methods, whereby one builds a core wavefunction consisting of a set amount of basis functions, which are optimized through multiple sweeps. Afterwards, additional basis functions are added one by one to "patch the holes" of the core basis. Thereby, one can to a large extend avoid redundant basis functions while retaining good convergence criteria.\\

One of the main issues of the Fully Correlated Gaussian method is its difficulty of describing the long ranged tail of the few-body wavefunctions. Thus, states with weak confinements or large spatial extensions are poorly approximated by a linear combination of Gaussians.
A possible solution is explicitly constructing the tail separately from the main body of the wavefunction by fitting an exponential function with a set amount of Gaussians. Thereby, part of the optimization procedure becomes matching the core and tail of the wavefunction. This solution should work well with the optimization scheme proposed above, as a set amount of Gaussian dedicated to describing the core and tail of the wavefunction can be optimized through a series of sweeps and subsequently be fit together by adding additional functions to the basis.\\

Another issue of the correlated Gaussian method is its lack of robustness, as various optimization algorithms may amplify numerical errors arising from taking matrix inverses. A poor choice of basis functions may result in almost singular matrices causing large numerical uncertainties. Optimization algorithms may attempt to exploit the numerical imprecision to reach lower energies than otherwise possible. A possible solution may be adding constraints to the problem, which ensures that the matrices remain invertible.

\section{Optimization Methods}
\begin{figure}[!h]
	\centering
	\includegraphics[width=1.1\columnwidth]{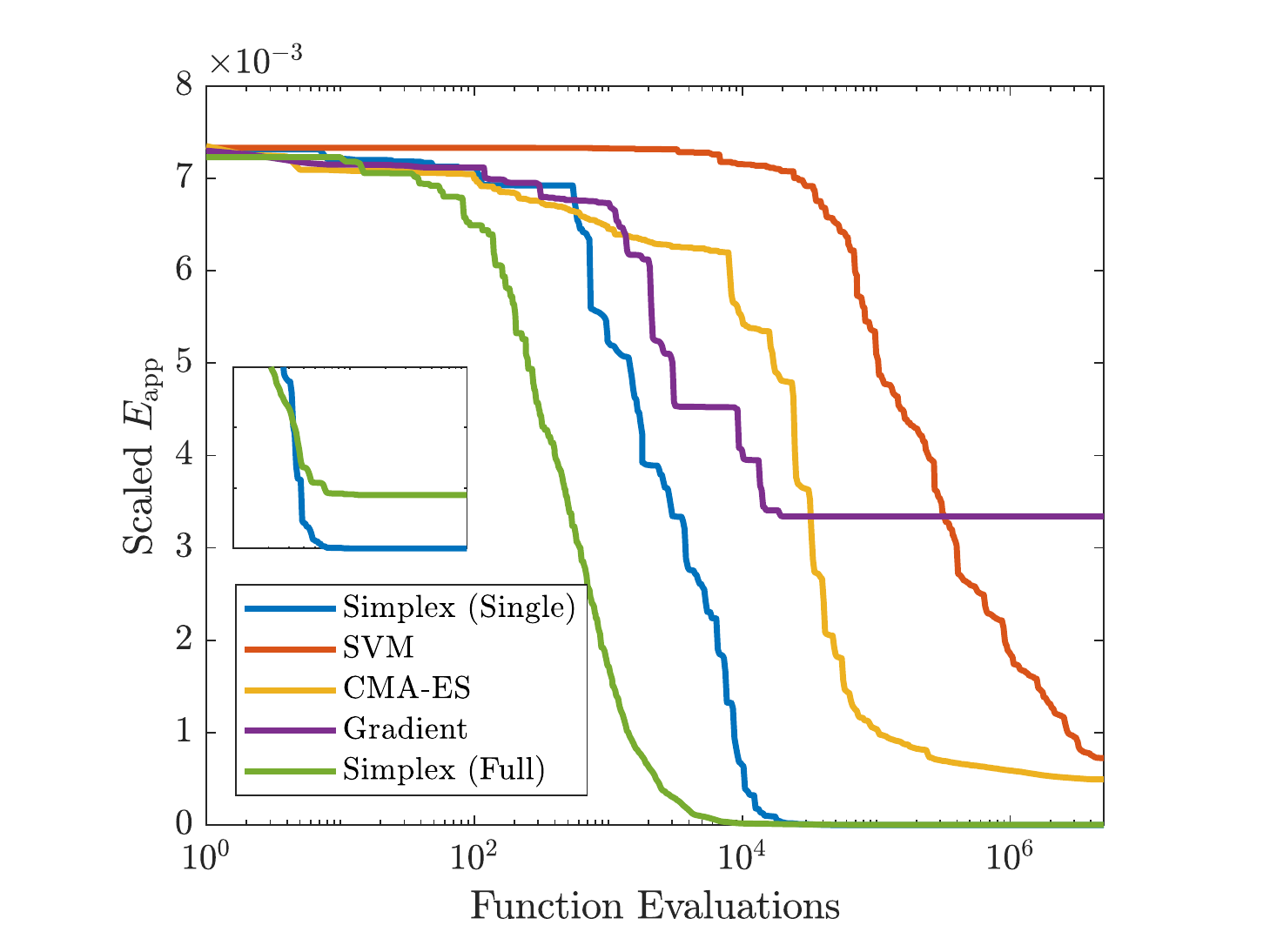} 
	\caption{First excited state of AAA system  calculated for moderate squeezing using various optimization methods. The calculated energy scaled according to lowest obtained value is plotted against the number of function evaluations performed by each method. The inset shows a zoomed in view of the results obtained with the simplex method.}
	\label{fig:Methods} 
\end{figure}

Various optimization algorithms for choosing the best linear combination of Gaussians in the variational search were tested. Figure \ref{fig:Methods} illustrates the rate of convergence of these algorithms when used for the three-body squeezing problem.  
The following briefly details the performance of each algorithm on the three-body squeezing problem:

\subsection{Stochastic Variational Method (SVM)}
In the stochastic variational methods the parameters of the Gaussians were randomly drawn from a distribution \cite{Varga1995}. In this case an exponential distribution was used. This method is very much a black box optimization method, as very little knowledge about the system is required. The algorithm performs well in cases where the optimization landscape is very complex or jagged, as this method cannot be stuck in a local minima due to the completely random nature of the parameter selection. Although the method is good at exploring the landscape, it converges very poorly to any minima, as there is no correlation between the parameters chosen.This can be seen in figure \ref{fig:Methods}, as the method steadily improves on the result without getting stuck, although the rate of convergence is very slow.
Thus, this method was primarily used to give a "warm start" to other algorithms.

\subsection{Covariance Matrix Adaptation Evolution Strategy (CMA-ES)}
Like SVM, the Covariance Matrix Adaptation Evolution Strategy samples parameters randomly from a distribution. However, this method contains an evolutionary element, whereby the best trial-parameters are used to perturb the shape and location of the sample-distribution \cite{Hansen2003}. Hence, the CMA-ES method converges much faster than SVM. The algorithm can also be tuned between exploration and exploitation of the optimization landscape by choosing the number of trial and parent parameters within each generation. While this method outperformed the stochastic variational method, it did not perform as well as the Nelder-Mead method.

\subsection{Nelder-Mead (Simplex)}
The Nelder-Mead method is a local optimization method, however, due to the final extension of the simplex, the method is capable of avoiding very narrow minima, which one may find in a jagged optimization landscape~\cite{Lagarias1998,Nocedal2006}. Thus, the Nelder-Mead method proved itself very robust when applied to the three-particle squeezing problem.
The robustness of the method was tested with regards to the number of basis function optimized at once. 
When optimizing the entire basis at once, the convergence would be very smooth and would generally require fewer function evaluations as illustrated in figure \ref{fig:Methods}. However, each function evaluation was much more time consuming, as the entire basis had to be updated rather than a single bass function. Another issue was a higher risk of getting stuck in local minima, which is shown in the inset of figure \ref{fig:Methods}. Meanwhile, optimizing only a single basis function at a time allowed the algorithm to "break out" of the local minima. Therefore, a combination of the two approaches was used. By optimizing two basis functions at a time, consistent convergence to the minimum was achieved without a large compromise of runtime.

\subsection{Gradient-Based Methods}
Gradient-based optimization methods generally converge faster than gradient-free methods such as Nelder-Mead~\cite{Nocedal2006}. Since all the correlated Gaussian matrix elements have analytical derivatives, utilizing the gradient for the optimization should increase the rate of convergence. However, due to the jaggedness of the optimization landscape, all gradient based approached attempted here converged to local minima, rather than the optimum. This can be seen in figure \ref{fig:Methods}, as the gradient-based method stops improving after only relatively few function evaluations. Gradient based methods were tested on optimizing a single basis function at a time versus optimizing the entire basis. Figure \ref{fig:Methods} displays the results of optimizing a single function, while considering the whole basis resulted in getting stuck after even fewer function evaluations.\\
A possible solution may be utilizing the second derivatives (Hessian) of the matrix elements. Optimization algorithms exploiting the Hessian to converge much faster than algorithms only considering the gradient. Furthermore, the additional information regarding the optimization landscape contained in the Hessian may help the algorithm stay clear of narrow, suboptimal minima. Luckily, all the matrix elements summarized in appendix \ref{chap:MatrixElem} have fairly simple second derivatives, whereby the elements of the Hessian can be computed rather efficiently. As discussed previously, adding constraints to the problem could prevent algorithms from exploiting numerical inaccuracies of the correlated Gaussian method. Many gradient-based optimization algorithm support constrained problems, however, one must derive the Jacobian and Hessian of the constraints as well. The interior point method \cite{Wachter2006,Nocedal2006} is an excellent algorithm for handling constrained, non-linear, high-dimensional optimization problems.


\begin{thebibliography}{63}
\bibitem [{Not()}]{NoteResonance}%
  \BibitemOpen
  \href@noop {} {}\bibinfo {note} {When mentioning resonance, we refer to the
  Feshbach resonance and the resulting infinite scattering length.}\BibitemShut
  {Stop}%
\end{thebibliography}
\end{document}